\documentclass{article}
\usepackage{epsfig}
\usepackage{amssymb}
\catcode`\@=11
\def\@normalsize{\@setsize\normalsize{15pt}\xiipt\@xiipt
\abovedisplayskip 14pt plus3pt minus3pt%
\belowdisplayskip \abovedisplayskip
\abovedisplayshortskip  \z@ plus3pt%
\belowdisplayshortskip  7pt plus3.5pt minus0pt}
\def\small{\@setsize\small{13.6pt}\xipt\@xipt
\abovedisplayskip 13pt plus3pt minus3pt%
\belowdisplayskip \abovedisplayskip
\abovedisplayshortskip  \z@ plus3pt%
\belowdisplayshortskip  7pt plus3.5pt minus0pt
\def\@listi{\parsep 4.5pt plus 2pt minus 1pt
            \itemsep \parsep
            \topsep 9pt plus 3pt minus 3pt}}
\def\underline#1{\relax\ifmmode\@@underline#1\else
        $\@@underline{\hbox{#1}}$\relax\fi}
\@twosidetrue \relax \catcode`@=12
\catcode`\@=11

\def\section{\@startsection{section}{1}{\z@}{3.5ex plus 1ex minus
   .2ex}{2.3ex plus .2ex}{\large\bf}}

\def\ps@headings{\def\@oddfoot{}\def\@evenfoot{}
\def\@oddhead{\hbox{}\hfill
        \makebox[.5\textwidth]{\raggedright\ignorespaces --\thepage{}--
        \hfill }}
\def\@evenhead{\@oddhead}
\def\subsectionmark##1{\markboth{##1}{}}
} \ps@headings \catcode`\@=12 \relax

\def\figcap{\section*{Figure Captions\markboth

        {FIGURECAPTIONS}{FIGURECAPTIONS}}\list
        {Fig. \arabic{enumi}:\hfill}{\settowidth\labelwidth{Fig. 999:}
        \leftmargin\labelwidth
        \advance\leftmargin\labelsep\usecounter{enumi}}}
 \relax
\def\tablecap{\section*{Table Captions\markboth
        {TABLECAPTIONS}{TABLECAPTIONS}}\list
        {Table \arabic{enumi}:\hfill}{\settowidth\labelwidth{Table 999:}
        \leftmargin\labelwidth
        \advance\leftmargin\labelsep\usecounter{enumi}}}
 \relax
\def\reflist{\section*{References\markboth
        {REFLIST}{REFLIST}}\list
        {[\arabic{enumi}]\hfill}{\settowidth\labelwidth{[999]}
        \leftmargin\labelwidth
        \advance\leftmargin\labelsep\usecounter{enumi}}}
 \relax
\catcode`\@=11
\def\marginnote#1{}
\newcount\hour
\newcount\minute
\newtoks\amorpm
\hour=\time\divide\hour by60 \minute=\time{\multiply\hour by60
\global\advance\minute by- \hour}
\edef\standardtime{{\ifnum\hour<12 \global\amorpm={am}%
    \else\global\amorpm={pm}\advance\hour by-12 \fi
    \ifnum\hour=0 \hour=12 \fi
    \number\hour:\ifnum\minute<100\fi\number\minute\the\amorpm}}
\edef\militarytime{\number\hour:\ifnum\minute<100\fi\number\minute}

\def\draftlabel#1{{\@bsphack\if@filesw {\let\thepage\relax
  \xdef\@gtempa{\write\@auxout{\string
    \newlabel{#1}{{\@currentlabel}{\thepage}}}}}\@gtempa
    \if@nobreak \ifvmode\nobreak\fi\fi\fi\@esphack}
     \gdef\@eqnlabel{#1}}
\def\@eqnlabel{}
\def\@vacuum{}
\def\draftmarginnote#1{\marginpar{\raggedright\scriptsize\tt#1}}
\def\draft{\oddsidemargin -.5truein
        \def\@oddfoot{\sl preliminary draft \hfil
        \rm\thepage\hfil\sl\today\quad\militarytime}

        \let\@evenfoot\@oddfoot \overfullrule 3pt
        \let\label=\draftlabel
        \let\marginnote=\draftmarginnote
\def\@eqnnum{(\theequation)\rlap{\kern\marginparsep\tt\@eqnlabel}%
\global\let\@eqnlabel\@vacuum}  }
\def\preprint{\twocolumn\sloppy\flushbottom\parindent 1em
        \leftmargini 2em\leftmarginv .5em\leftmarginvi .5em
        \oddsidemargin -.5in    \evensidemargin -.5in
        \columnsep 15mm \footheight 0pt
        \textwidth 250mmin      \topmargin  -.4in
        \headheight 12pt \topskip .4in
        \textheight 175mm
        \footskip 0pt
\def\@oddhead{\thepage\hfil\addtocounter{page}{1}\thepage}
        \let\@evenhead\@oddhead \def\@oddfoot{} \def\@evenfoot{}
}
\def\titlepage{\@restonecolfalse\if@twocolumn\@restonecoltrue\onecolumn
     \else \newpage \fi \thispagestyle{empty}\c@page\z@
        \def\thefootnote{\fnsymbol{footnote}} }
\def\endtitlepage{\if@restonecol\twocolumn \else  \fi
        \def\thefootnote{\arabic{footnote}}
        \setcounter{footnote}{0}}  
\catcode`@=12 \relax

\def\ps@headings{\def\@oddfoot{}\def\@evenfoot{}
\def\@oddhead{\hbox{}\hfill
        \makebox[.5\textwidth]{\raggedright\ignorespaces --\thepage{}--
        \hfill }}
\def\@evenhead{\@oddhead}
\def\subsectionmark##1{\markboth{##1}{}}
} \ps@headings \relax
\newcommand{\newc}{\newcommand}

\newc{\ra}{\rightarrow}
\newc{\lra}{\leftrightarrow}
\newc{\beq}{\begin{equation}}
\newc{\be}{\begin{equation}}
\newc{\eeq}{\end{equation}}
\newc{\ee}{\end{equation}}
\newc{\bea}{\begin{eqnarray}}
\newc{\eea}{\end{eqnarray}}
\def\eps{\epsilon}

\newc{\ome}{\omega}
\newc{\ba}{\begin{eqnarray}}
 \newc{\ea}{\end{eqnarray}}

\begin{document}
\def\firstpage#1#2#3#4#5#6{
\begin{titlepage}
\nopagebreak
\title{\begin{flushright}
        \vspace*{-0.8in}
\end{flushright}
\vfill {#3}}
\author{\large #4 \\[1.0cm] #5}
\maketitle \vskip -7mm \nopagebreak
\begin{abstract}
{\noindent #6}
\end{abstract}
\vfill
\begin{flushleft}
\rule{16.1cm}{0.2mm}\\[-3mm]

\end{flushleft}
\thispagestyle{empty}
\end{titlepage}}

\def\simlt{\stackrel{<}{{}_\sim}}
\def\simgt{\stackrel{>}{{}_\sim}}
\date{}
\firstpage{3118}{IC/95/34} {\large\bf  Inverted neutrino mass
hierarchies from $U(1)$ symmetries.}
 {G.K. Leontaris$^a$ A. Psallidas$^a$ and N.D. Vlachos$^b$}
{\normalsize\sl $^a$ Theoretical Physics Division, Ioannina University,
GR-45110 Ioannina, Greece\\
\normalsize\sl $^b$ Dept. of Theoretical Physics, University
of Thessaloniki, GR-54124 Thessaloniki, Greece\\ [2.5mm]
 }
{Motivated by effective low energy models of string origin, we
discuss the neutrino masses and mixing within the context of the
Minimal Supersymmetric Standard Model supplemented by a $U(1)$
anomalous family symmetry and additional Higgs singlet fields
charged under this extra $U(1)$. In particular, we  interpret the
solar and atmospheric neutrino data assuming that there are only
three left-handed neutrinos which acquire Majorana masses via a
lepton number violating dimension-five operator. We derive the
general form of the charged lepton and neutrino mass matrices when
two different pairs of singlet Higgs fields develop non--zero vacuum
expectation values and show how the resulting neutrino textures are related to
approximate lepton flavor symmetries. We perform a numerical
analysis for one particular case and obtain solutions for masses and
mixing angles, consistent with  experimental data. }

\vskip 3truecm

\newpage

\section{Introduction}
Analysis of the atmospheric and solar neutrino oscillation data
\cite{Fukuda:2000np,rdata,Smirnov:2004ju} imply tiny neutrino
squared mass differences and  large mixing.  Moreover, Yukawa
couplings related to neutrino masses are highly suppressed compared
to those for quarks and charged leptons.

Various  authors~\cite{mt}-\cite{Grimus:2004hf}, have claimed that
the neutrino mass matrix structure and the (almost) maximal
$\nu_{\mu}-\nu_{\tau}$ mixing -as the atmospheric data suggest-
could be interpreted in terms of a symmetry beyond the standard
model. Motivated by the fact that the majority of string models
constructed so far include several (possibly anomalous) additional
$U(1)$'s,  we consider that
this picture may indicate an underlying structure of the
mass matrix determined by such a $U(1)$ broken at some high scale
$M$. An apparently different approach has shown that the neutrino
data can  be interpreted using a  restricted  class of mass
matrices~\cite{Frampton:2002qc} where at most two of its entries
vanish.  It is likely that these zeros can be naturally generated in
the context of a symmetry principle obeyed by the neutrino Yukawa
couplings.

Light Majorana neutrino masses in the range below the $eV$ scale -as
experiments suggest- can be obtained either through a see-saw
mechanism (in the presence of right handed-neutrinos), or from the
dimension-five non-renormalizable operator $(L\,H)^2/M$, where $L,H$
are the lepton and Higgs doublets respectively, while $M$ is an
appropriate large scale. In  previous
work\cite{Leontaris:1999wf,Leontarisxx} we have suggested that the
Minimal Supersymmetric  Standard Model (MSSM) extended by a single
$U(1)$ anomalous family symmetry spontaneously broken by non--zero
vacuum expectation values (vevs) of a  pair of singlet fields $\Phi,
\bar\Phi$ with $U(1)$ charges $Q=\pm 1$ can provide acceptable
masses and large mixing. It was found that such a symmetry retains
the above mentioned dimension-five operator which provides Majorana
masses to the left-handed neutrino states. Assuming symmetric lepton
mass matrices, it was shown that the neutrino sector respects an
$L_e-L_{\mu}-L_{\tau}$ symmetry, implying inverse hierarchical
neutrino mass spectrum,  large solar ($\theta_{12}$) and atmospheric
($\theta_{23}$) mixing angles whilst $\theta_{13}=0$ and degeneracy
between the first two neutrino mass eigenstates.  In view of these
shortcomings, it was suggested that second order effects -possibly
originating from additional singlet field vevs- might lift the mass
degeneracy and reconcile the neutrino data accurately. In the
present work, we explore in detail the structure of the lepton mass
matrices in the presence of non-renormalizable contributions
originating from more than one singlet Higgs pairs.  This extension
is motivated by the fact that string models usually predict a large
number of singlet fields charged differently under the extra $U(1)$
symmetry. We start with a systematic study of the mass matrix
structures obtained from non-renormalizable contributions coming
from $\Phi,\bar \Phi$ singlet vevs only and establish the
consistency of the resulting neutrino texture forms with flavor
symmetries of the type $L_e\pm (L_{\mu}\mp L_{\tau})$. Next, we
consider non-renormalizable, hierarchically lesser contributions
from additional singlets, and work out the particular case
$L_f=L_e-L_{\mu}-L_{\tau}$ in detail. We show that contributions
from a proper second Higgs singlet pair, generate additional mass
entries, breaking `softly' the symmetry $L_f$. We find that -under
natural assumptions for the undetermined Yukawa coefficients- it is
possible to correlate the resulting neutrino mass matrices with
particular texture zeros discussed in~\cite{Frampton:2002qc} and
give a consistent set of models which are in agreement with recent
data.

The paper is organized as follows. In section 2, we develop a useful
texture form of the neutrino mass matrix in terms of the eigenmasses
and mixing angles and give a brief description of the constraints
implied by recent oscillation data. In section 3, we present the
extension of the model~\cite{Leontarisxx} with the inclusion of the
second singlet Higgs pair and derive the relevant charged lepton and
Majorana neutrino mass matrices. In section 4, we perform a test
calculation and give solutions consistent with current neutrino
data. Our conclusions are presented in section 5.

\section{Neutrino mass matrix constraints and oscillation data}

In this work, we assume that the neutrino data can be interpreted in
terms of a Majorana mass matrix of the left-handed neutrino
components. There are at most nine independent parameters in this
matrix, while experiments can only measure two squared mass
differences, three angles, one CP-phase and the double beta decay
parameter. Thus, the neutrino mass matrix cannot be fully determined
by the present experiments. A general neutrino mass matrix will look as follows:%
\begin{equation}
M^{\nu }=U_{n}M_{d}U_{n}^{\intercal }  \label{m1}
\end{equation}%
where $M_{d}$ is the diagonalized Majorana neutrino mass matrix
\begin{equation}
M_{d}=\left[
\begin{array}{ccc}
m_{1} & 0 & 0 \\
0 & m_{2} & 0 \\
0 & 0 & m_{3}%
\end{array}%
\right]  \label{d1}
\end{equation}%
and $U_{n}$ is the diagonalizing matrix parametrized in terms of three
angles ($c_{ij}\equiv \cos\theta_{ij}, s_{ij}\equiv \sin\theta_{ij}$)
\begin{equation}
U_{n}\ =\left[
\begin{array}{ccc}
c_{12}{\ \,}c_{13} & c_{13}{\ \,}s_{12} & s_{13}{\,}e^{-i\delta} \\
{\ -}c_{23}{\ \,}s_{12}{\ -}c_{12}{\ \,}s_{13}{\ \,}s_{23}{\,}e^{i\delta} & c_{12}{\ \,}%
c_{23}{\ -}s_{12}{\ \,}s_{13}{\ \,}s_{23}{\,}e^{i\delta} & c_{13}{\ \,}s_{23} \\
{\ -}c_{12}{\ \,}c_{23}{\ \,}s_{13}{\,}e^{i\delta}{\ +}s_{12}{\ s}_{23} & {\ -}c_{23}{\ \,}%
s_{12}{\ \,}s_{13}{\ -}c_{12}{\ \,}s_{23}{\,}e^{i\delta} & c_{13}{\ \,}c_{23}%
\end{array}%
\right] {\ \quad \cdot }  \label{rotmat}
\end{equation}%
We observe that the matrix elements of (\ref{m1}) can be written as inner
products
\begin{equation}
M^{\nu }=\left[
\begin{array}{ccc}
\overrightarrow{m}\cdot \overrightarrow{\vartheta_{1}} & \overrightarrow{m}%
\cdot \overrightarrow{\vartheta_{2}} & \overrightarrow{m}\cdot
\overrightarrow{\vartheta_{3}} \\
\overrightarrow{m}\cdot \overrightarrow{\vartheta_{2}} & \overrightarrow{m}%
\cdot \overrightarrow{\vartheta_{4}} & \overrightarrow{m}\cdot
\overrightarrow{\vartheta_{5}} \\
\overrightarrow{m}\cdot \overrightarrow{\vartheta_{3}} & \overrightarrow{m}%
\cdot \overrightarrow{\vartheta_{5}} & \overrightarrow{m}\cdot
\overrightarrow{\vartheta_{6}}%
\end{array}%
\right] \quad \cdot \label{text1}
\end{equation}%
where $\overrightarrow{m}=\left( m_{1},m_{2},m_{3}\right) $ is the
vector of the mass eigenstates (complex in general) and the vectors
$\overrightarrow{\vartheta_{i}}$ are functions of the mixing angles
$\theta _{12}$, $\theta _{13}$, $\theta _{23}$ and the phase
$\delta$.
\begin{eqnarray*}
\overrightarrow{\vartheta _{1}} &=&\left[ {c}_{12}^{2}{c}_{13}^{2},{c}%
_{13}^{2}{s}_{12}^{2},{s}_{13}^{2}{\,}e^{-2i\delta }\right]  \\
\overrightarrow{\vartheta _{2}} &=&\left[ -c_{12}\,c_{13}\,\left(
c_{23}\,s_{12}+c_{12}\,s_{13}\,s_{23}{\,}e^{i\delta }\right)
,c_{13}~s_{12}\,\left(
c_{12}\,c_{23}-s_{12}\,s_{13}\,s_{23}~e^{i\delta
}\right) ,c_{13}\,s_{13}\,s_{23}~e^{-i\delta }\right]  \\
\overrightarrow{\vartheta _{3}} &=&\left[ c_{12}~c_{13}\,\left(
s_{12}\,s_{23}-c_{12}\,c_{23}\,s_{13}~e^{-i\delta }\right)
,-c_{13}\,s_{12}\,\left( c_{23}\,s_{12}\,s_{13}~e^{i\delta
}+c_{12}\,s_{23}\right) ,c_{13}\,c_{23}\,s_{13}e^{-i\delta }\right]  \\
\overrightarrow{\vartheta _{4}} &=&\left[ {\left(
c_{23}\,s_{12}+c_{12}\,s_{13}\,s_{23}~e^{i\delta }\right)
}^{2},{\left(
c_{12}\,c_{23}-s_{12}\,s_{13}\,s_{23}~e^{i\delta }\right) }^{2},{c}%
_{13}^{2}\,{s}_{23}^{2}\right]  \\
\overrightarrow{\vartheta _{5}} &=&\left[ \left(
c_{12}\,c_{23}\,s_{13}e^{i\delta }-s_{12}\,s_{23}\right)\left(
c_{23}\,s_{12}+c_{12}\,s_{13}\,s_{23}e^{i\delta }\right),\right.\\
&& \left. \left( c_{23}\,s_{12}\,s_{13}e^{i\delta
}+c_{12}\,s_{23}\right) \left(
s_{12}\,s_{13}\,s_{23}e^{i\delta }-c_{12}\,c_{23}\right),{c}_{13}^{2}\,c_{23}\,s_{23}\right]  \\
\overrightarrow{\vartheta _{6}} &=&\left[ {\left(
c_{12}\,c_{23}\,s_{13}~e^{i\delta }-s_{12}\,s_{23}\right)
}^{2},{\left(
c_{23}\,s_{12}\,s_{13}~e^{i\delta }+c_{12}\,s_{23}\right) }^{2},{c}_{13}^{2}~%
{c}_{23}^{2}\right] \cdot
\end{eqnarray*}
Pairs of $\overrightarrow{\vartheta _{i}}$ vectors can be checked to
be in general linearly independent thus, in a three dimensional
space, $\overrightarrow{m}$ can be made orthogonal to at most one
pair. In this case, (\ref{m1}) assumes a texture form i.e.,
$\overrightarrow{m}\cdot \overrightarrow{\vartheta _{i}}=0$,
$~\overrightarrow{m}\cdot \overrightarrow{\vartheta _{j}}=0$ and we
get a constraint on the actual values of the neutrino masses
\begin{equation}
\overrightarrow{m}=m_{0}\,\overrightarrow{\vartheta _{i}}\times
\overrightarrow{\vartheta _{j}}\   \label{const1}
\end{equation}%
where $m_{0}$ is a (complex) mass parameter which characterizes the
neutrino mass scale. The parameter $|m_{0}|$ takes values in a
closed interval which can be determined by requiring the moduli of
the mass eigenstates to satisfy the experimental bounds
\begin{eqnarray}
5.4~10^{-5} &\leq &|m_{2}|^{2}-|m_{1}|^{2}\leq 9.5~10^{-5}  \label{bound1} \\
1.4~10^{-3} &\leq &|m_{3}|^{2}-|m_{2}|^{2}\leq 3.7~10^{-3}\ \cdot
\nonumber
\end{eqnarray}%
Experimentally allowed values for the elements of $M^{\nu }$ can be
determined as follows: We first form the dimensionless ratio
\begin{equation}
14.73\leq
\frac{|m_{3}|^{2}-|m_{2}|^{2}}{|m_{2}|^{2}-|m_{1}|^{2}}\leq 68.51\
\label{bound2}
\end{equation}%
which depends only on the values of the mixing angles $\theta
_{12}$, $\theta _{13}$, $\theta _{23}$. Then, we look for values of
the angles within the experimental bounds
\begin{eqnarray}
0.23 &\leq &\sin ^{2}\theta _{12}\leq 0.39   \nonumber \\
0.31 &\leq &\sin ^{2}\theta _{23}\leq 0.72  \label{bound3}\\
0.00 &\leq &\sin ^{2}\theta _{13}\leq 0.054  \nonumber
\end{eqnarray}%
satisfying the ratio constraint (\ref{bound2}), which are depicted in figures 1-4.
These angle values can now be used to construct the matrix $M^{\nu }$.
Note that, the experimentally allowed values for $\theta _{13}$ are
close to zero, also those for $\theta _{23}$ are close to $\pi /4$.
In the limit where $\theta _{13}=0$ and $\delta=0$ we get the
additional constraints $\overrightarrow{\vartheta _{2}}+
\overrightarrow{\vartheta _{3}}=0$ and $\overrightarrow{\vartheta _{5}}\cdot%
\left( \overrightarrow{\vartheta _{4}}\times \overrightarrow{\vartheta _{6}}%
\right) =0$. If we further require that $\theta_{23}=\pi /4$ we
get in addition
$\overrightarrow{\vartheta_{4}}=\overrightarrow{\vartheta _{6}}$.
As an example, we assume that $\overrightarrow{m}\cdot \overrightarrow{%
\vartheta _{4}}=0$ and $\overrightarrow{m}\cdot \overrightarrow{\vartheta
_{6}}=0$, so that the Majorana neutrino mass matrix takes the form
\begin{equation}
M^{\nu }=\left[
\begin{array}{ccc}
\overrightarrow{m}\cdot \overrightarrow{\vartheta _{1}} & \overrightarrow{m}%
\cdot \overrightarrow{\vartheta _{2}} & \overrightarrow{m}\cdot
\overrightarrow{\vartheta _{3}} \\
\overrightarrow{m}\cdot \overrightarrow{\vartheta _{2}} & 0 &
\overrightarrow{m}\cdot \overrightarrow{\vartheta _{5}} \\
\overrightarrow{m}\cdot \overrightarrow{\vartheta _{3}} & \overrightarrow{m}%
\cdot \overrightarrow{\vartheta _{5}} & 0%
\end{array}%
\right] \ \cdot   \label{ex1}
\end{equation}%
If this texture is to be compatible with experimental data, we
must expect that $ M^{\nu }_{12}= M^{\nu }_{21}\approx - M^{\nu
}_{13}$ and $ M^{\nu }_{23}\ll 1$.

The above formulation reveals another interesting property of the
matrix (\ref{ex1}). If the mixing $\nu_{\mu}-\nu_{\tau}$ is exactly
maximal, i.e., if $\theta_{23}=\frac{\pi}4$, then the texture
(\ref{ex1}) automatically implies degeneracy of the two eigenvalues.
One finds
\ba
m_1\equiv m_2&=&-m_0 \cos^2\theta_{13}\\
m_3&=&\,m_0 \left(1+e^{2\imath \delta}\sin^{2}\theta_{13}\right) \ea
Thus, exactly maximal mixing in this case is not allowed since it
would imply $\Delta\,m_{sol.}^2=0$, in clear disagreement with the
experimental data.

\section{Description of the Model}

In what follows we shall assume that there is no CP violation in the
neutrino sector, and all fermionic mass eigenstates are real, but
not necessarily positive. Our intention is to interpret the
oscillation neutrino data using only the Standard Model fermion
spectrum (without  right handed neutrinos) and  two pairs of singlet
Higgs fields with appropriate $U(1)$ charges. The particles,
together with their $U(1)_X$ charge notation are presented in Table
1. The $U(1)_X$ charges of $\Phi,\bar\Phi$ are taken to be $\pm 1$,
while the charges of the other representations will be fixed from
low energy physics considerations.

 In the absence of the right handed neutrinos, the neutrino
masses arise from  the lepton number violating ($\Delta L=2$)
operator \cite{Weinberg:sa,Barbieri:1980hc}
\ba
 \frac{y_{\nu}^{a \beta} }{M }
  (\bar{L_{a}^{c}}^{i} H^{j} \epsilon_{ji})
  (H^{l} L_{\beta}^{k}  \epsilon_{lk}) &
  \equiv& \frac{y_{\nu}^{a \beta}v^2 }{M }
   \bar\nu_{La}^c\nu_{L\beta}\label{neff}
  \ea
Here, $y_{\nu}^{a \beta}$ is an effective Yukawa coupling depending
on the details of the theory, $v=\langle H\rangle$ is the Higgs
doublet vev which is of the order of the electroweak scale and $M$
stands for a large scale $M\sim 10^{13-14}$ GeV, which could be
identified with the effective gravity scale in theories with large
extra dimensions  obtained in the context of Type I string models.
\footnote{ We also notice that models with intermediate scales of
the above order can be obtained in the context of D-brane scenarios,
see for example~\cite{Gioutsos:2005uw}.}

To start with, we fist review the construction with the introduction
of the first singlet Higgs field pair only, i.e. $\Phi,\,\bar\Phi$
with charges $\pm 1$ and define for convenience the ratios $\lambda
=\frac{\langle\Phi\rangle}M$ and $\bar\lambda
=\frac{\langle\bar\Phi\rangle}M$. We choose the fermion $U(1)_X$
charges of the Standard Model particles presented in Table 1, so
that only the third generation tree-level couplings  are present at
the tree level potential
\ba
{\cal W}_{tree} &=& y_{33}^u Q_3 U^c_3 H_2 + y_{33}^dQ_3 D^c_3 H_1
               +y_{33}^e L_{3} E^c_3 H_1
\label{sup2tree}
\ea
with $y_{u,d,e}$ being the order-one Yukawa couplings. The two
lighter generations get masses through non renormalizable
Yukawa-type interactions. Thus, for the up, down quarks and charged
leptons these are
\ba
{\cal W}_{n.r.}^{(1)}\propto Q_i U_j^c H_2 \varepsilon^{C^u_{ij}}+
 Q_i D_j^c H_1 \varepsilon^{C^d_{ij}}
 + L_i E_j^c H_1 \varepsilon^{C^l_{ij}} \nonumber
\ea
where, $C_{ij}^a$,   ($a=u,d,l$) are appropriate integer powers
depending on the $U(1)_X$-charge of the corresponding Yukawa term
and $\varepsilon$ is defined as follows
\ba
\varepsilon^{k}=\left\{
\begin{array}{ll}
\lambda^{k}&\mbox{if}\ k=[k]<0\\
\bar{\lambda}^{k}&\mbox{if}\ k=[k]>0\\
0&\mbox{if}\ k\ne[k]
\end{array}
\right. \label{edef}
\ea
Neutrino masses are generated by the operator (\ref{neff}), suppressed
by the appropriate powers of the parameter $\varepsilon$.
\begin{table}
\begin{center}
\begin{tabular}{|cc|cc|}
\hline
Fermion&$U(1)_X$-charge&Higgs&$U(1)_X$-charge\\
\hline
$Q_i(3,2,\frac 16)$&$q_i$&$H_1(1,2,-\frac 12)$&$h_1$\\
$D^c_i(\bar 3,1,\frac 13)$&$d_i$&$H_2(1,2,\frac 12)$&$h_2$\\
$U^c_i(\bar 3,1,-\frac 23)$&$u_i$&--------------------&------\\
$L_i(1,2,-\frac 12)$&$\ell_i$&$\Phi,\,\bar\Phi(1,1,0)$&$\pm 1$\\
$E^c_i(1,1,1)$&$e_i$&$\chi,\,\bar\chi(1,1,0)$&$\pm Q_{\chi}$\\
\hline
\end{tabular}
\end{center}
\caption{\label{uxa}$U(1)_X$-charge assignments for MSSM fields.
The $U(1)_X$ charges of the  singlet pair fields $\Phi$ and
$\bar\Phi$ are taken to be $\pm 1$ while the corresponding charges of
$\chi,\,\bar\chi$ will be fixed by phenomenological requirements.
}
\end{table}

We assume symmetric mass matrices and take into account the fact
that the third generation appears at tree-level. This way, the charged lepton
$U(1)_X$--charge matrix takes the form
\ba
C_{e}=\left(\begin{array}{ccc}
n'&\frac{m'+n'}{2}&\frac{n'}{2}\\
\frac{m'+n'}{2}&{m'}&\frac{m'}{2}\\
\frac{n'}{2}&\frac{m'}{2}&0\\
\end{array}\right)\cdot
\label{lep1}
\ea
Here, $m',n'$ are taken to be
integers\cite{Leontaris:1999wf,Leontarisxx} and we have defined
$\frac{n'}2=\ell_1-\ell_3$ and $\frac{m'}2=\ell_2-\ell_3$, where $\ell_i$ are the
$U(1)_X$-charges of the leptons.
Given the form of the
operator (\ref{neff}) we find that the $U(1)_X$-charge entries for
the light Majorana neutrino mass matrix takes the form
\ba
C_{\nu}= \left(\begin{array}{ccc}
n'+{\cal A}&\frac{m'+n'}{2}+{\cal A}&\frac{n'}{2}+{\cal A}\\
\frac{m'+n'}{2}+{\cal A}&{m'}+{\cal A}&\frac{m'}{2}+{\cal A}\\
\frac{n'}{2}+{\cal A}&\frac{m'}{2}+{\cal A}&{\cal A}\\
\end{array}\right)
\ea where ${\cal A}$ is another constant, expressed in terms of
$U(1)$ fermion charges, ${\cal A}=2 (\ell_3+h_2)$. Thus, the
neutrino $U(1)_X$--charge entries differ from the corresponding
charged leptonic entries by the constant ${\cal A}$.

\subsection{Lepton Mass Matrices}

 We start now our investigation of the $M^{e}$ and $M^{\nu}$  mass
 matrices considering first the implications of singlets $\Phi,\bar\Phi$.
A mass entry $m_{ij}^e$ of $M^e$ is non-zero whenever a power
$C_{ij}^e$ of the expansion parameter $\eps$ in (\ref{edef})
matches the charge of the corresponding entry of (\ref{lep1}).
Since the charges with respect to $U(1)_X$ of the latter are
$Q_X=\pm 1$, an entry $m_{ij}^e$ is non-zero whenever the
corresponding  entry in (\ref{lep1}) is also integer. Since
$m',n'$ are integers, in order to obtain non-zero entries in the
Majorana matrix too, {\it the parameter ${\cal A}$ has to be
either integer or half-integer.} The case where ${\cal A}$ is
integer implies that the charged lepton and neutrino mass matrices
are proportional $M^e\sim \eps^{\cal A} M^{\nu}$ (up to Yukawa
coefficients) and it is hard to obtain large mixing. Thus, we will
assume that ${\cal A}$ is half integer and we will analyze this
case in the sequel.

As far as $m',n'$ parameters are concerned, we distinguish the
following cases with respect to neutrino mass matrix: $(i)$: if we
take $m'$ even and $n'$ odd, i.e. $m'=2p$ and $n'=2q+1$, (where $p,q$
are integers), then only the elements $m^{\nu}_{12}=m^{\nu}_{21}$,
 $m^{\nu}_{13}=m^{\nu}_{31}$ are non-zero in the neutrino mass matrix.
 In this case, the neutrino mass matrix respects an $L_f=L_e-L_{\mu}-L_{\tau}$
flavor symmetry, (where $L_{e,\mu,\tau}$ are the lepton flavor
numbers of the three generations), implying bimaximal
($\theta_{12},\theta_{23}$) mixing, inverted neutrino mass hierarchy
and, in particular, degeneracy of the two first neutrino eigenstates
$m^{\nu}_1=m^{\nu}_2$ while $m^{\nu}_3=0$. The corresponding charged
lepton mass matrix, however, does not conserve $L_f$. ($ii$): if
both $m'$ and $n'$ are odd, then $m^{\nu}_{13}=m^{\nu}_{31}\ne 0$ as
well as $m^{\nu}_{23}=m^{\nu}_{32}\ne 0$, while all other neutrino
mass entries are zero. We note that this neutrino texture exhibits
an $L_e+L_{\mu}-L_{\tau}$ symmetry. ($iii$): in the case $m'=2p+1$,
$n'=2q$, the non-zero elements are $m^{\nu}_{12}=m^{\nu}_{21}\ne 0$,
$m^{\nu}_{23}=m^{\nu}_{32}\ne 0$, while the symmetry is
$L_e-L_{\mu}+L_{\tau}$, and finally $(iv)$: if both parameters are
even integers, $m'=2p$, $n'=2q$, all the entries of $C_{\nu}$ become
half-integers and the neutrino matrix is zero in this case,
respecting thus an $L_e+L_{\mu}+L_{\tau}$ symmetry. On the contrary,
all $M^e$ mass matrix elements are non-zero, filled-in by
non-renormalizable terms $\eps^{C_{ij}^e}$. All four cases are
summarized in Table \ref{ytab2}.
\begin{table}[!ht]
\centering
\begin{tabular}{|c|c|c|c|c|}
\hline%
$m'$&$n'$&${\cal A}$&$m^{\nu}_{ij}\ne 0$&$M^{\nu}$-Symmetry\\
\hline%
 $2\,p$& $2\,q+1$ & $r+1/2$& $m^{\nu}_{12},m^{\nu}_{21},m^{\nu}_{13},m^{\nu}_{31}$&$L_e-L_{\mu}-L_{\tau}$ \\
\hline%
$2\,p+1$& $2\,q+1$ &$r+1/2$& $m^{\nu}_{13},m^{\nu}_{31},m^{\nu}_{23},m^{\nu}_{32}$&$L_e+L_{\mu}-L_{\tau}$ \\
\hline%
$2\,p+1$& $2\,q$ &$r+1/2$&$m^{\nu}_{12},m^{\nu}_{21},m^{\nu}_{23},m^{\nu}_{32}$& $L_e-L_{\mu}+L_{\tau}$ \\
\hline  $2\,p$& $2\,q$ &$r+1/2$&none& $L_e+L_{\mu}+L_{\tau}$ \\
\hline
\end{tabular}
\caption{\label{ytab2} The symmetries of the Majorana neutrino mass textures
in the presence of only one singlet Higgs pair $\Phi,\bar\Phi$ with $U(1)_X$
charges $\pm 1$, for ${\cal A}=r+1/2$ and
the four distinct cases of the integers $m',n'$ (see text for details).}
\end{table}


The above analysis shows that, if only one singlet acquires non-zero
vev, the neutrino data cannot be accommodated. Assuming
however, that the low energy spectrum of  the higher theory contains
various neutral singlet scalars -as it is the case in string
constructions-, we expect that contributions from an appropriate
second singlet vev will  suffice to reconcile the predictions of the
modified mass matrices with the experimental data.

From inspection of the mixing angles in the above four cases, we
infer that  ($ii$) and ($iii$) are unlikely to interpret the
neutrino data. On the contrary, for case ($i$) small contributions
may modify the neutrino masses and mixing and lead to acceptable
results. Finally, as already noted, neutrino masses are all zero in
case ($iv$) at this level, however, a second singlet with proper
$U(1)_X$ charge can in principle generate also a viable $M^{\nu}$
matrix.

 We will analyze in some detail case ($i$), i.e.,
  $n' = \mbox{odd} \ ,\ m' = \mbox{even}$ and show that
 a viable set of lepton mass matrices and mixing
 naturally arise. If we write $m'=2\,p$, $n'=2\,q+1$, ${\cal A}=r+1/2$ with $p,q,r$
integers, the $U(1)_X$-charge entries of the charged-lepton mass
matrix are~\footnote{The corresponding quark matrices have been
analyzed in ref~\cite{Leontarisxx}.}
\ba
C_{e}=\left(\begin{array}{ccc}
2q+1&p+q+\frac{1}{2}&q+\frac{1}{2}\\
p+q+\frac{1}{2}&2p&p\\
q+\frac{1}{2}&p&0\\
\end{array}\right)\cdot
\label{lep2}
\ea
The corresponding charge entries for the neutrino mass matrix are
\ba
C_{\nu}=\left(\begin{array}{ccc}
2q+r+\frac 32&p+q+r+1&q+r+1\\
p+q+r+1&2p+r+\frac 12&p+r+\frac 12\\
q+r+1&p+r+\frac{1}{2}&r+\frac 12\\
\end{array}\right)\cdot
\label{Cnu2}
\ea
The  parameters $m',$\,$n'$,$\,{\cal A}$ (or $p,q,r$
respectively), determine also the $U(1)_X$-charges of the SM
particles of Table 1. A systematic search shows that a natural set
of $U(1)_X$ charges is obtained~\cite{Leontarisxx} for $m'=2,
n'=7, {\cal A}=-\frac 52$. (For the lepton doublets in particular,
choosing for example $h_2=\frac 14$ we find
$\ell_1=2,\ell_2=-\frac 12,\ell_3=-\frac 32$). In this case, the
leptonic charge entries become
\ba
C_{e}=\left(\begin{array}{ccc}
7&\frac 92&\frac 72\\
\frac 92&2&1\\
\frac 72&1&0\\
\end{array}\right),
&
 C_{\nu}=\left(\begin{array}{ccc}
\frac 92&2&1\\
2&-\frac 12&-\frac 32\\
1&-\frac{3}{2}&-\frac 52\\
\end{array}\right)\cdot
\label{cnu2} \ea When only the singlet Higgs pair $\Phi,\bar\Phi$
with $Q_{X}=\pm 1$ obtain vevs, as already pointed out, only the
integer charge mass entries are filled in. In this case, there are
more than two zeros in the neutrino mass matrix thus, the
experimental data can not be accurately interpreted
~\cite{Frampton:2002qc}. Indeed, one finds $\Delta m^2_{atm}=\Delta
m_{23}^2$ and $ \Delta m^2_{sol}=\Delta m_{12}^2=0$, i.e., the
first two neutrino mass eigenstates are degenerate. Furthermore, the
leptonic mixing matrix $U_l^0 = V_l^\dagger V_{\nu}$ implies that
the solar neutrino mixing angle is maximal, a situation disfavored
by recent data.

As noted above, at this stage, the neutrino mass matrix respects a
symmetry of the form $L_f=L_e-L_{\mu}-L_{\tau}$, where
$L_{e,\mu,\tau}$ are the lepton flavor numbers of the three
generations. We could consider this structure of the neutrino matrix
as a starting point and consider additional secondary effects which
break the $L_f$ symmetry  softly and lead to a texture form
consistent with experimental data. Indeed, the above drawbacks could
be cured if additional non-zero entries are generated by additional
effects. As noted in the introduction, in a string model, the low
energy spectrum contains more than one Higgs singlets. A new singlet
pair $\chi,\bar\chi$ with appropriate $U(1)_X$-charges might develop
a vev, (provided that the flatness conditions are satisfied) so that
additional mass entries could be filled-in by additional
non-renormalizable terms.

Checking the neutrino charge entries in the case under
consideration(\ref{cnu2}), we conclude that in order to have a viable
neutrino mass matrix by symmetry principles, we need to generate
non-zero values at least for the $\{23\}, \{32\}$ and $\{11\}$
elements, so that a matrix of the  form (\ref{ex1}) of section 2
could be obtained. For example, in order to get a non-zero $\{32\}$
entry, the charge of the second singlet should satisfy $s\,Q_{\chi}
-\frac 32=0$, where $s$ is integer. This implies
$Q_{\chi}=\frac{3}{2s}\ra \{\frac 32,\frac 34,\frac 12,\frac
38,\dots\}$. A reasonable choice of $U(1)_X$ charges of this extra
singlet Higgs pair is $Q_{\chi}=\frac{3}{2}$ and
$Q_{\bar\chi}=-\frac{3}{2}$.

Let us now assume  that the new singlets $\chi,\bar\chi$, obtain
vevs and denote $\eta =\frac{\langle\chi\rangle}M$,
$\bar\eta=\frac{\langle\bar\chi\rangle}M$. In analogy to
(\ref{edef}) we define $\vartheta^{k}=\eta^{k}$ {if} $k=[k]<0$,
$\vartheta^{k}=\bar{\eta}^{k}$ {if} $k=[k]>0$ and $\vartheta^{k}=0$
otherwise. Then, for $Q_{\chi}=\frac 32$, a mass matrix element
$m_{ij}$ with charge entry $N_{ij}+\frac 12$, (where $N_{ij}$ the
integer part) is \ba
m_{ij}&=&\sum_{\ell=\cdots-1,0,1\cdots}y_{ij}^{\ell}\;
\eps^{N_{ij}-3\ell-1}\,\vartheta^{2\ell+1}\label{NRt} \ea where
$y_{ij}^{\ell}$ are order-one Yukawa coefficients. Let $p=1,q=3$ so
that $m'=2,n'=7$ as in case (\ref{cnu2}). Then, \ba
m^e_{12}\;=\;y_{12}^1\;\bar\lambda^3\bar\eta+y_{12}^2\;\bar\eta^3+\cdots,&
m^e_{13}\;=\;y_{13}^1\bar\lambda^2\bar\eta+\cdots\nonumber \ea and
similarly for the rest of the mass matrix elements. We should point
out however, that in a realistic string construction, due to
additional (discrete) symmetries -remnants of the original string
symmetry and various selection rules-, several non-renormalizable
terms in the sum (\ref{NRt})  will vanish.

To construct the mass matrices, we assume that $\lambda,\bar\lambda$
and $\eta,\bar\eta$ are in the perturbative region, so we  retain
non-renormalizable terms only up to fourth order. Ignoring for
simplicity the coefficients $y_{ij}$, the mass matrices are written
in terms of $\lambda,\bar\lambda$ and $\eta,\bar\eta$ as follows.
For the charged leptons we find \ba
   M_e \approx m_0^e\left(
  \begin{array}{ccc}
          \bar\lambda^{7}                 &  \bar\lambda^3\bar\eta+\bar\eta^3    & \bar\lambda^2\bar\eta +\lambda\bar\eta^3\\
  \bar\lambda^3\bar\eta+\bar\eta^3        &  \bar\lambda^{2}+\bar\eta^2\lambda   & \bar\lambda+\lambda^2\bar\eta^2   \\
  \bar\lambda^2\bar\eta +\lambda\bar\eta^3& \bar\lambda +\lambda^2\bar\eta^2                & 1 \\
  \end{array}
  \right),
  \label{palep3}
\ea
and for the neutrinos~\footnote{If only contributions
of $\eta,\bar\eta$ are taken into account, the neutrino mass matrix exhibits
an $L_{\mu}-L_{\tau}$ symmetry. }
\ba
M_{\nu}\approx m_0\left(\begin{array}{ccc}
 \bar\lambda^3\bar\eta+\bar\eta^3  &  -\bar\lambda^{2}+\bar\eta^2\lambda   &\bar\lambda+\lambda^2\bar\eta^2 \\
 -\bar\lambda^2+\bar\eta^2\lambda  &  \bar\lambda\eta+\lambda^2\bar\eta &  \eta\\
 \bar\lambda+\lambda^2\bar\eta^2   &  \eta &\lambda\eta\\
\end{array}\right)\cdot
\label{nu2A}
\ea
Thus, in the presence of two singlet Higgs pairs, in principle,
all the entries of the charged lepton and neutrino mass matrices become
non-zero. We may assume however, hierarchies between the vevs
$\lambda,\bar\lambda,\eta,\bar\eta$, and  derive approximate texture
zero mass matrices~\cite{Frampton:2002qc} in the limiting cases
where some of the vacuum expectation values are taken to be zero or
negligible compared to others. Note  that the corresponding charged
lepton mass matrices are not diagonal for these neutrino textures,
thus a straightforward comparison with the results of
~\cite{Frampton:2002qc}  is not possible. We will see however, that
off-diagonal entries and mixing effects of the charged lepton sector
in our models are not substantial, thus the neutrino mixing angles
receive only small contributions from the charged lepton mass
matrix.  Thus, if we assume that $\lambda\ra 0$  , $\eta \ll
\bar\eta $ and $\bar\eta <\bar\lambda$, we obtain the analogue of
texture (\ref{ex1})
\ba
M_{\nu }=m_{0}\left[
\begin{array}{ccc}
\overline{\eta }^{3} & -\overline{\lambda }^{2} &  \overline{\lambda } \\
-\overline{\lambda }^{2} & \sim 0 & \eta \\
 \overline{\lambda } & \eta & 0%
\end{array}%
\right] \ \cdot  \label{m20}
\ea

If we now take $\bar\lambda \ll 1$, and ignoring fourth order
terms, if $\lambda <\bar\eta$ we can approximate
\ba
M_{\nu}=m_0\left(\begin{array}{ccc}
\bar\eta^3  &  \bar\eta^2\lambda   &\sim 0 \\
\bar\eta^2\lambda  &  \sim 0 &  \eta\\
 \sim 0   &  \eta &\lambda\eta\\
\end{array}\right) \cdot
\label{nu2A2}
\ea
 If on the other hand we take $\lambda >\bar\eta$, we get
 the approximate form
\ba
M_{\nu}=m_0\left(\begin{array}{ccc}
 \sim 0  &  \bar\eta^2\lambda   &\sim 0\\
 \bar\eta^2\lambda  &  \lambda^2\bar\eta &  \eta\\
\sim 0&  \eta &\lambda\eta\\
\end{array}\right) \cdot
\label{nu2B1}
\ea
For $\bar\eta \ll 1$, while assuming $\bar\lambda^2\ll 1$ we
obtain
\ba
M_{\nu}=m_0\left(\begin{array}{ccc}
 \sim 0  &  \sim 0   &\bar\lambda\\
 \sim 0 &  \eta \bar \lambda &  \eta\\
\bar\lambda&  \eta &\lambda\eta\\
\end{array}\right)\cdot
\label{nu2A1}
\ea
All textures (\ref{m20}-\ref{nu2A1}) are  of the form discussed in
\cite{Frampton:2002qc}.  In the next section, we are going to
further analyze one specific case out of the many possible that may
occur. Similar results can be obtained for the other cases too.

We give now a brief account of case $(iv)$. A viable case arises
if we choose $n'=8,m'=2$ and ${\cal A}=-\frac 92$. The choice
$h_2=-\frac 14$ for the Higgs $U(1)_X$-charge, implies also a
natural set of lepton charges in this case,
$\ell_1=2,\ell_2=-1,\ell_3=-2$.  Furthermore, for
$Q_{\chi,\bar\chi}=\pm\frac 12$, we obtain the matrices
\ba
M^{e}=m_0^e\left(\begin{array}{ccc}
\bar\lambda^8 & \bar\lambda^5 & \bar\lambda^4 \\
\bar\lambda^5 & \bar\lambda^2 & \bar\lambda  \\
\bar\lambda^4 & \bar\lambda   &    1       \\
\end{array}\right),&
M_{\nu}=m_0\left(\begin{array}{ccc}
 \bar\eta^7 & \bar\eta &\eta   \\
 \bar\eta   & \eta^5   &\eta^7 \\
 \eta       & \eta^7   &\eta^9 \\
\end{array}\right)\cdot
\label{caseiv} \ea which also lead to inverted neutrino mass
hierarchy and large mixing for $\eta\approx \bar\eta$.

\section{Numerical Analysis}

In this section we will calculate the mass spectrum and the leptonic
mixing angles for the texture (\ref{m20}) which is of the type
$\overrightarrow{m}=m_{0}\,\overrightarrow{\vartheta_{4}}\times
\overrightarrow{\vartheta_{6}}$ of section 2. A texture of this type
leads to inverted hierarchy values for $m_{1}$ $%
m_{2}$ $m_{3}$, i.e., $m_{1}\sim $ $m_{2}>m_{3}$. Note also that
the sign of $m_{2}$ always comes out negative.  In the
approximation $\eta\ll\bar\eta$ and $\lambda \ll 1$ the model
matrix~(\ref{nu2A}) becomes
\begin{equation}
M^{\nu }=m_{0}\left[
\begin{array}{ccc}
\overline{\eta }^{3} & -\overline{\lambda }^{2}
& \xi \overline{\lambda }
\\
-\overline{\lambda }^{2} & 0 & \eta \\
\xi \overline{\lambda } & \eta & 0%
\end{array}%
\right] \ \cdot  \label{m2}
\end{equation}%
where $\xi$ is a Yukawa parameter of order one.  In the limit
$\eta,\,\bar\eta \ra 0$ we have the simpler one singlet case
discussed previously. This is the case where the neutrino sector
respects the symmetry $L_f=L_e-L_{\mu}-L_{\tau}$ discussed
extensively in the literature ~\cite{Petcov:1982ya,Petcov:2004rk}
which implies bimaximal ($\theta_{21}, \theta_{23}$) mixing. Mixing
effects of the charged lepton matrix and the second singlet field
result to a ``soft breaking'' of the quantum number $L_f$.
Using the analysis of section 2, it is straightforward to generate
numerical values for $\chi, $ $\xi, $ $\overline{\lambda },$ that
satisfy the bounds (\ref{bound1},\ref{bound2},\ref{bound3}). It is
to be noted however, that only a subset of the so generated values
can satisfy the constraints implied by the charged-leptons mass
matrix.

We now turn on to the charged lepton mass matrix.  Since the
leptonic mass matrix has fewer degrees of freedom than the
corresponding neutrino matrix, additional Yukawa parameters have to
be introduced.
The minimal number required in order to reach a solution is three ($%
m_{0}^{e}, $ $b,$ $d$ ). This way (using in the same approximation
$\eta\ll\bar\eta, \lambda \ll 1$), the leptonic mass matrix becomes
\begin{equation}
M^{e}=m_{0}^{e}\left[
\begin{array}{ccc}
\overline{\lambda }^{7} & b\,\overline{\eta }^{3} & 0 \\
b\,\overline{\eta }^{3} & \overline{\lambda }^{2} & d\,\overline{\lambda } \\
0 & d\,\overline{\lambda } & 1%
\end{array}%
\right] \ \cdot  \label{m3}
\end{equation}%
If we are to be working in the perturbative region, we must require
the values of $\overline{\eta }$ and $\overline{\lambda }$ to be
smaller than one, while the values of the constants $b$ and $d$ to
be at the most of order one. The trace of the matrix $M^{e}$ equals
the sum of the lepton masses, thus, we expect that the value of the
constant $m_{0}^{e}$ to be around the value of the tau mass. Upon
further numerical investigation, solutions that satisfy our criteria
were found. In table~\ref{tab1}  we present a selected subset. The
columns $(2-5)$ of this table present the numerical values of
$m_{0}^{e}$, $b$, $d$,  $\overline{\eta }$ and $\overline{\lambda
}$.
All these numbers correspond to lepton masses $m_{e}=-0.511~\mathrm{MeV}$ $%
m_{\mu }=105.66~\mathrm{MeV}$ and $m_{\tau}=1777.05~\mathrm{MeV}$.
The matrix $M^{e}$ can be diagonalized by means of an orthogonal
matrix $U_{l}$ so that
$$\mathrm{diagonal}\left[ m_{e},m_{\mu },m_{\tau }\right]
=U_{l}M^{e}U_{l}^{\intercal }$$ The strong hierarchy of the leptonic
masses $|m_{e}|\ll m_{\mu }<m_{\tau }$ and the structure of the
charged lepton mass matrix obtained, put strong constraints on the
elements of $U_{l}$. Since $M_{11}^{e}$ and $M_{12}^{e}$ are
negligible, the eigenvectors corresponding to the eigenvalues
$m_{\tau }$ and $m_{\mu }$ must be of the form $e_{3}=\left[ \sim
0,\sim 0,\sim 1\right] $ and $e_{2}=\left[ \sim 0,\sim 1,\sim
0\right] $ by orthogonality. Orthogonality then implies that the
third eigenvector $e_{1}=\left[ \sim 1,\sim 0,\sim 0\right] $.
Numerics confirm this statement. Indeed, choosing from Table \ref{tab1}
the first solution we get
\[
U_{l}=\left[
\begin{array}{ccc}
0.993337 & -0.112408 & 0.0254002 \\
0.115231 & 0.965657 & -0.23287 \\
0.00164859 & 0.234245 & 0.972176%
\end{array}%
\right]
\]%
while for the eleventh solution we get
\[
U_{l}=\left[
\begin{array}{ccc}
0.997154 & -0.0753675 & 0.00201336 \\
0.0753943 & 0.996752 & -0.0283198 \\
0.000127579 & 0.028391 & 0.999597%
\end{array}%
\right] \ \cdot
\]%

\begin{center}
\begin{table}
\begin{center}
\begin{tabular}{|r|r|r|r|r|r|r|r|}
\hline
$n_{0}$ & $m_{0}^{e}$ & $b$ & $d$ & $\overline{\eta }$ & $\overline{\lambda }
$&$\xi $ & $\eta $ \\ \hline
1 & 1685.27 & 0.106904 & 0.662742 & 0.410916 & 0.341056 &0.241615& 0.072256\\ \hline
2 & 1685.33 & 0.114839 & 0.662618 & 0.401166 & 0.340994& 0.24267 & 0.064982\\ \hline
3 & 1694.54 & 0.109728 & 0.643608 & 0.39885 & 0.33211  & 0.239861& 0.063127\\ \hline
4 & 1700.30 & 0.09143 & 0.630442 & 0.418435 & 0.326475& 0.234507& 0.074418 \\ \hline
5 & 1717.55 & 0.088453 & 0.583331 & 0.407698 & 0.309176 & 0.229562& 0.068114\\ \hline
6 & 1735.09 & 0.090929 & 0.518201 & 0.390316 & 0.290874& 0.223921 & 0.058959 \\ \hline
7 & 1755.26 & 0.116885 & 0.401999 & 0.346965 & 0.268733 & 0.217379& 0.040692\\ \hline
8 & 1759.96 & 0.124525 & 0.362853 & 0.337367 & 0.263384& 0.215189& 0.037264 \\ \hline
9 & 1772.21 & 0.103441 & 0.203602 & 0.353084 & 0.24901  & 0.205992& 0.044141\\ \hline
10 & 1775.19 & 0.110105 & 0.128032 & 0.344582 & 0.245416 & 0.204041& 0.039409\\ \hline
11 & 1775.70 & 0.127301 & 0.109161 & 0.328114 & 0.24479& 0.204962& 0.033917  \\ \hline
\end{tabular}
\end{center}
\caption{\label{tab1} The values of the parameters for 11 selected cases satisfying
experimental data for lepton masses and mixing angles.
$m_{0}^{e}$, $b$, $d$, $\overline{\eta }$ and  $\overline{\lambda }$ are determined from
the charged lepton mass matrix. The remaining $\xi $ and $\eta $ are fixed by $m_{\nu}$.}
\end{table}
\end{center}
Thus, $U_{l}$ being close to the unit matrix in all cases, transfers
only a small mixing to the neutrino sector. The neutrino mixing
matrix is given by
\[
M^{\nu }=U_{l}^{\intercal }M_{\nu }U_{l}
\]%
and can be diagonalized by means of an orthogonal matrix which is
to be identified with $U_{n}$ (\ref{rotmat}). Columns $(6,7)$ of
Table \ref{tab1} show the values of the remaining parameters
$\eta, \xi$ fixed by the neutrino mass matrix. From this
identification the values of the mixing angles $\theta _{12}$,
$\theta _{13}$, $\theta _{23}$ can be calculated.

Table \ref{tab3} summarizes the results found for the neutrino
mass scale parameter $m_0$ and the mixing angles $\theta_{ij}$
which are consistent with the recent data. In the first two columns,
we show the minimum and maximum bounds on $m_0$ which defines the
overall neutrino mass scale in (\ref{m3}). The remaining columns,
summarize the values for the three mixing angles.
\begin{center}
\begin{table}
\begin{center}
\begin{tabular}{|r|r|r|r|r|r|r|}
\hline $n_{0}$ &$ m_{0\min }$ & $m_{0\max }$  & $\sin
^{2}\theta _{12}$ & $\sin ^{2}\theta _{13}$ & $\sin ^{2}\theta _{23}$ \\
\hline 1 & 0.377324 & 0.421945 & 0.37482 & 0.053167 & 0.549865
\\\hline 2 &  0.235305 & 0.312101 & 0.389747 & 0.044787 & 0.555663
\\ \hline
3 & 0.227853 & 0.302218 & 0.379206 & 0.042058 &
0.552185 \\ \hline 4 & 0.259091 & 0.343651 & 0.342147 & 0.055435 &
0.539129 \\ \hline 5 & 0.280139 & 0.371568 & 0.329176 & 0.046816 &
0.531798 \\ \hline 6 & 0.301829 & 0.400338 & 0.325867 & 0.036055 &
0.521448 \\ \hline 7 &
0.367913 & 0.487989  & 0.358503 & 0.019655 & 0.503853 \\
\hline 8 & 0.385085 & 0.510765& 0.365593 & 0.017262 & 0.496102 \\
\hline 9 & 0.670619 & 0.753771 & 0.315931 & 0.022377 & 0.453056 \\
\hline
10 & 0.377407 & 0.500582  & 0.327493 & 0.018809 & 0.437599 \\
\hline 11 & 0.404905 & 0.537054 & 0.353621 & 0.015066 & 0.436352
\\ \hline
\end{tabular}
\end{center} \caption{\label{tab3}The range of the neutrino mass
scale $m_0$ defined in (\ref{m3}) is shown in the first two columns.
The values of $\sin^2\theta_{ij}$ for the eleven cases discussed
in the text are also presented in the next three columns.}
\end{table}
\end{center}
We finally discuss in brief  the prediction for the neutrinoless
double-beta ($\beta\beta_{0\nu}$) decay. Operator (\ref{neff})
violates lepton number by two units, allowing thus the
$\beta\beta_{0\nu}$ process $(A,Z)\ra (A,Z+2)+2 e^-$. The parameter relevant to experiments
for this process is
\ba
m_{ee}&=&\sum_{i}U_{e\,i}^2m_{\nu_i}\label{mee}
\ea
while the experimental constraint is ${m_{ee}}^{exp}
\le [0.55-1.10]\,$eV\cite{Strumia:2005tc}.  The exact predictions for all
eleven cases for ${m_0=m_0}_{max}$ are compatible with the experimental bounds and are presented in Table
\ref{tab4}.

We proceed with some remarks on the solutions obtained. We first observe
that all vevs are in the perturbative region, while $\eta \ll
\bar\eta \sim\bar\lambda $ as already assumed. It can be checked that for all
solutions  $m_{11}^{\nu}\approx m_{23}^{\nu}$, since $\bar\eta^3\sim
\eta$. Furthermore, we observe that around solution 8 we get the approximate relation
$\theta_{12}+\theta_{13}\approx \theta_{23}$. This  could be
related to a recent conjecture about  quark-lepton
complementarity (QLC) \cite{Minakata:2004xt} which states that
$\theta_{12}^{c}\approx \theta_{13}^{\nu}$ and
$\theta_{12}^{c}+\theta_{12}^{\nu}\approx \frac{\pi}4$  where
$\theta_{12}^{c}$ is the Cabbibo angle.

\begin{table}
\begin{center}
\begin{equation}
\begin{tabular}{|r|r|r|r|r|}
\hline
${m_{0}}_{max}$ & $m_{\nu_1}$ & $m_{\nu_2}$ & $m_{\nu_3}$ & $m_{ee}$ \\
\hline 0.421945 & 0.0677911& $-$0.0672912& 0.0287763 & 0.0178\\
\hline 0.312101 & 0.0496165& $-$0.0486496& 0.0191827& 0.0117\\
\hline 0.302218 & 0.0460616&  $-$0.0450185& 0.0181325 & 0.0118\\
\hline 0.343651 &0.0525161&  $-$0.0516036& 0.0242644 & 0.0173\\
\hline 0.371568  & 0.0516527&  $-$0.0507248& 0.024252& 0.0182\\
\hline 0.400338  & 0.049542&  $-$0.0485738& 0.0228372 & 0.0178\\
\hline 0.487989 & 0.0501541&  $-$0.0491979& 0.0194268 & 0.0146\\
\hline 0.510765 & 0.0502146&  $-$0.0492598& 0.0186574& 0.0139\\
\hline 0.753771& 0.0695465&  $-$0.0690545& 0.0326878& 0.0259\\
\hline 0.500582  & 0.0446707&  $-$0.0435943& 0.0194047 & 0.0158\\
\hline 0.537054 & 0.0464765& $-$0.0454431& 0.0179377 & 0.0140\\
\hline
\end{tabular}
\end{equation}\end{center}
\caption{\label{tab4} Neutrino masses and the double beta decay
parameter $m_{ee}=\sum  U_{ei}^2m_i$ for the maximum
value of the neutrino mass scale parameter $m_0$ of (\ref{m3}).
All masses are expressed in $eV$ units.}
\end{table}
\bigskip
\begin{figure}[tbp]
\begin{center}
\epsfxsize=11cm\epsfysize=9cm\epsffile{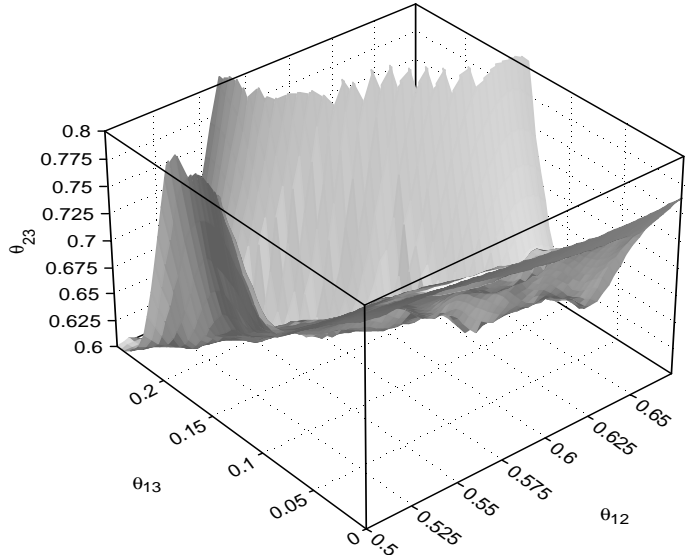}
\end{center}
\caption{The allowed region for the three neutrino mixing angles}
\end{figure}
\begin{figure}[tbp]
\begin{center}
\epsfxsize=11cm\epsfysize=9cm\epsffile{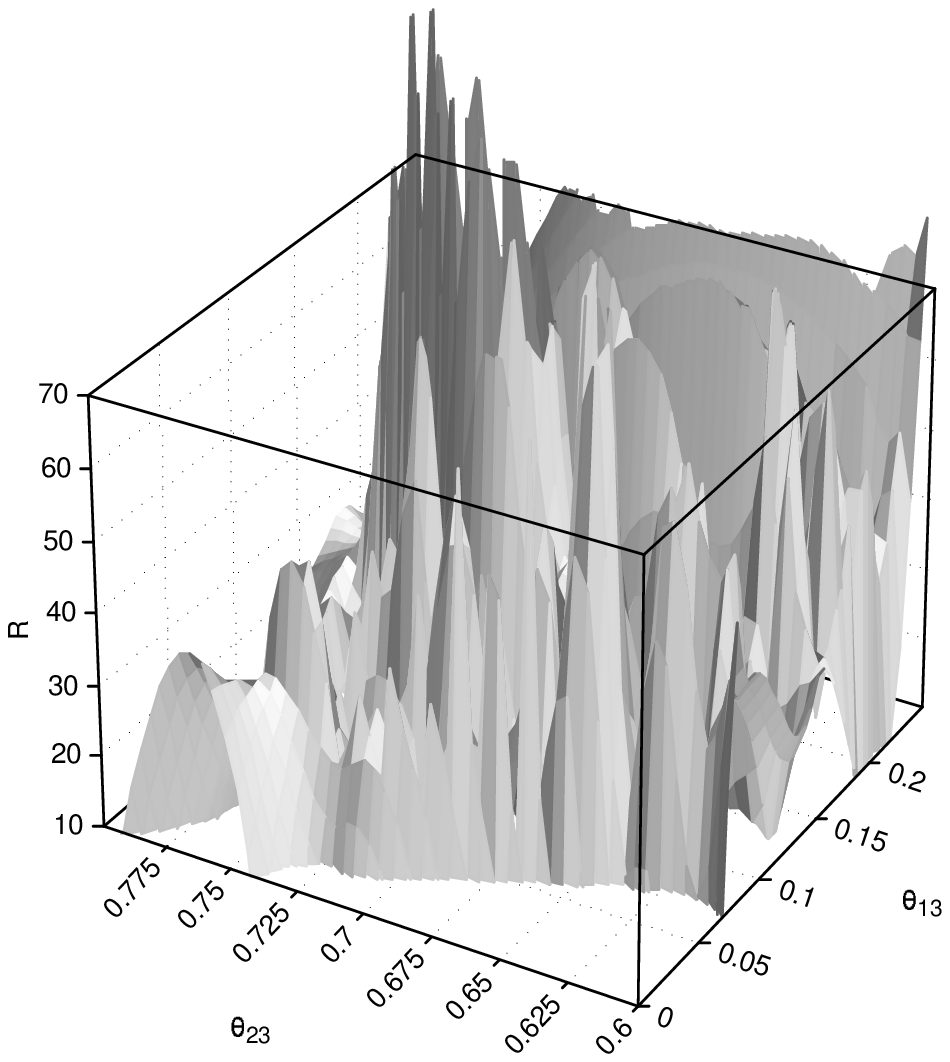}
\end{center}
\caption{The dimensionless neutrino mass ratio (\ref{bound2}) as a function of the corresponding angles}
\end{figure}
\begin{figure}[tbp]
\begin{center}
\epsfxsize=11cm\epsfysize=9cm\epsffile{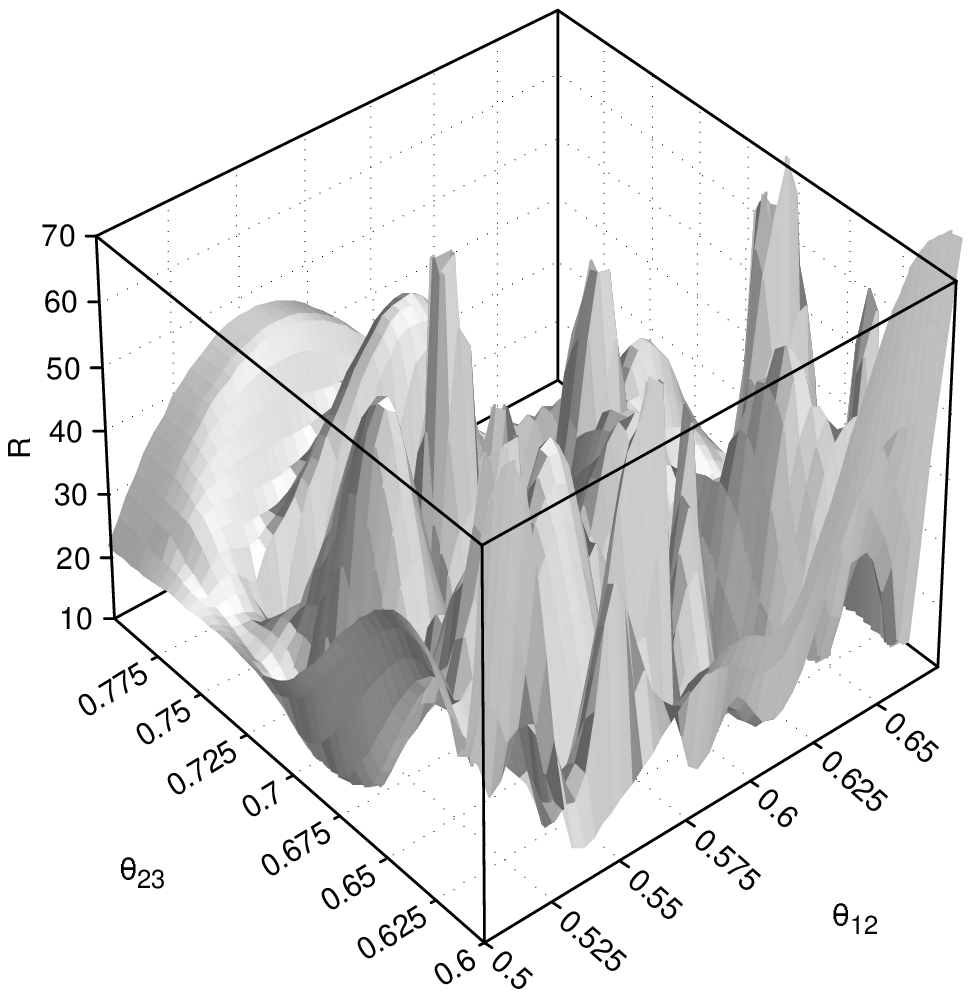}
\end{center}
\caption{The dimensionless neutrino mass ratio (\ref{bound2}) as a function of the corresponding angles}
\end{figure}
\begin{figure}[tbp]
\begin{center}
\epsfxsize=11cm\epsfysize=9cm\epsffile{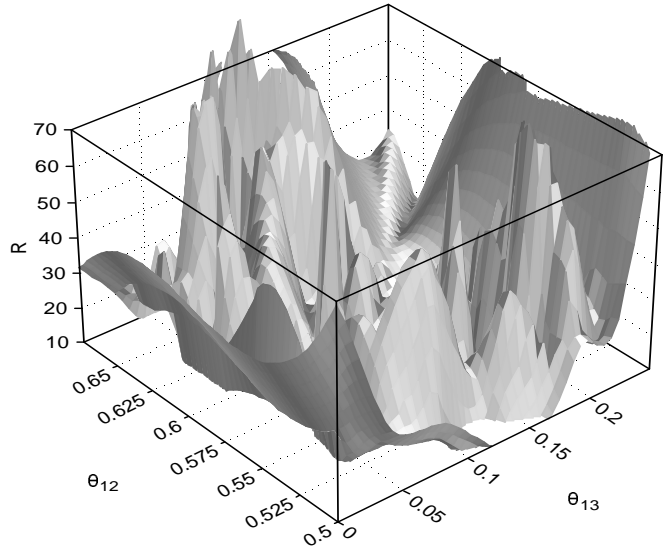}
\end{center}
\caption{The dimensionless neutrino mass ratio (\ref{bound2}) as a function of the corresponding angles}
\end{figure}
\section{Conclusions}

In this work we have explored the possibility of deriving viable
neutrino mass matrix textures capable of interpreting the recent
neutrino data for masses and mixing. This was realized  in the
context of the Minimal Supersymmetric Standard Model (MSSM) extended by an Abelian
symmetry with a minimal fermion content  and two additional pairs of
singlet Higgs fields. We have developed a useful texture formalism
where the matrix elements are represented as inner products of the
neutrino mass eigenstate vector $\vec m=(m_1,m_2,m_3)$ with six
vectors $\vec\vartheta_{i}$ which are functions of the mixing angles.
Texture matrices having two vanishing
elements~\cite{Frampton:2002qc}, $\vec m\cdot \vec\vartheta_{i}=0$,
$\vec m\cdot \vec\vartheta_{j}=0$ imply a neutrino eigenmass vector
of the form $\vec{m}=m_0\vec\vartheta_i\times\vec\vartheta_j$. We have
attempted to relate these structures with textures obtained from
symmetry principles in the context of the above proposed   extended
MSSM model. It was found that Yukawa mass terms, generated when the
two singlet Higgs pairs with appropriate $U(1)$ charges obtain vevs,
can generate to a good approximation, several texture zero forms.
We have analyzed the particular case where the corresponding neutrino
mass texture form respects the $L_f=L_e-L_{\mu}-L_{\tau}$ lepton
flavor symmetry implying bimaximal neutrino mixing when only one
singlet Higgs pair receives a non-zero vev. Contributions  from the second singlet Higgs pair vev
break the $L_f$ symmetry and generate new mass entries leading to acceptable
neutrino masses and mixing. A numerical analysis, confirms that for natural values of the --yet undetermined
by the theory-- Yukawa couplings, there exist solutions satisfying all experimental constraints.

\section*{Acknowledgements}

\par

\noindent {\it This research was supported by the program  `HERAKLEITOS'  of the
Operational Program for Education and Initial Vocational Training of
the Hellenic Ministry of Education under the 3rd Community Support
Framework and the European Social Fund, and, by the European Union under contract MRTN-CT-2004-503369.}

\end{document}